\documentclass[apj]{emulateapj}

\slugcomment{ApJ, in press}
\shorttitle{MCMC Catalog}
\shortauthors{Veras and Ford}

\begin{document}
\title{Secular Orbital Dynamics of Hierarchical Two Planet Systems}
\author{Dimitri Veras\altaffilmark{1}, Eric B. Ford\altaffilmark{1}}
\altaffiltext{1}{Astronomy Department, University of Florida, 211 Bryant Space Sciences Center, Gainesville, FL 32111, USA}
\email{veras@astro.ufl.edu}
\begin{abstract}
The discovery of multi-planet extrasolar systems has kindled 
interest in using their orbital evolution
as a probe of planet formation.  Accurate descriptions of
planetary orbits identify systems which 
could hide additional planets or be in a special dynamical
state, and inform targeted follow-up observations.  
We combine published radial velocity data with Markov Chain 
Monte Carlo analyses in order 
to obtain an {\it ensemble} of masses, semimajor axes, eccentricities 
and orbital angles for each of 5 dynamically active multi-planet systems: HD 11964, HD 38529, 
HD 108874, HD 168443, and HD 190360.
We dynamically evolve these systems using 52,000 long-term N-body integrations
that sample the full range of possible line-of-sight 
and relative inclinations, and we report on the system
stability, secular evolution and the extent of the resonant interactions.  
We find that planetary orbits in hierarchical systems exhibit complex dynamics
and can become highly
eccentric and maybe significantly inclined.  Additionally we
incorporate the effects of general relativity in the long-term simulations
and demonstrate that can qualitatively affect the dynamics
of some systems with high relative inclinations.  The simulations quantify
the likelihood of different dynamical regimes for each system 
and highlight the dangers of restricting simulation phase space 
to a single set of initial conditions or coplanar orbits.

\end{abstract}


\section{Introduction}

Currently, over 30 multi-planet exosystems are each known
to include 2-5 known planets. The orbital architectures
and formation scenarios for these systems have become
subjects of numerous investigations.
The recent discovery of a fifth planet
around 55 Cnc \citep{fisetal2008} has prompted a flurry
of follow-up studies \citep{gayetal2008,rayetal2008a,jietal2009} 
that aim to better describe the dynamics and 
evolution of that system.
Despite sparse data, the directly-imaged triple system 
HR 8799 \citep{maretal2008} has been subject to intense scrutiny 
\citep{fabmur2008,clomal2009,fuketal2009,gozmig2009,lafetal2009,reietal2009,sudetal2009}.
These explorations
demonstrate that major open questions
regarding multi-planet systems remain, including: What is the 
timescale for instability in these systems?  Could
they admit additional, currently undetectable planets?  How tightly 
``packed'' are such systems (e.g. \citealt*{rayetal2009})?

These questions can be addressed by investigating the orbital
evolution of individual planetary systems.  How do
the orbital eccentricities and semimajor axes of planets change with
time?  How does additional physics (e.g. tidal forces
and general relativity, henceforth referred to as GR) affect the 
orbits of short-period planets and
indirectly other planets in the system?  Theoretical investigations
addressing these questions should account for the uncertainties 
in orbital parameters obtained 
from observations.  In order to study dynamics, we need to 
establish initial conditions.  Unfortunately, the accuracy of 
such initial conditions are limited
due to measurement uncertainties and degeneracies
inherent to the radial velocity (RV) exoplanet discovery technique.
Occasionally, the best-fit RV data yields parameters which indicate
that the timescale for such a system to undergo instability 
is much less than the age of the system (e.g., in less than
$10^5$ yr, for HD 82943; \citealt*{mayetal2004,fabmur2008}).  These
short timescales (relative to system lifetimes)
suggest that the best-fit orbital model is unlikely and motivate investigations
to find the plausible and stable solutions.
Further, both $i_{rel}$ and $i_{LOS}$  for planets
in most multi-planet systems have not yet been measured.
Noteworthy exceptions include a recent 
estimate of the relative inclination ($i_{rel}$)
of GJ 876 planets \citep{beasei2009} and recent estimates for the
line-of-sight inclinations ($i_{LOS}$) of the planets in HR 8799 
\citep{lafetal2009}.
  
In an effort to remedy some of these issues and better describe
system properties, investigators have been developing techniques to 
model RV data and to describe their orbital evolution.  \cite{gozetal2005} 
and \cite{gozetal2008} utilize stability constraints and 
optimization techniques partly 
derived from the MEGNO (Mean Exponential Growth Factor 
for Nearby Orbits) method \citep{cinsim2000}.  The advantages 
to MEGNO include simultaneous multi-planet fitting and 
efficient identification of
quasi-periodic or irregular (chaotic) motion.  A 
disadvantage is that the arbitrary choice 
of the ``penalty'' parameter determining this timescale
prevents a rigorous interpretation of the results.
Some authors \citep{ford2005,ford2006,gregory2007a,gregory2007b}
use Markov Chain Monte Carlo (MCMC) techniques
to generate a sample of initial conditions from the
posterior probability distribution.
The strength of these techniques relies on
rigorous Bayesian calculation and interpretations,
and a weakness is that stability must be individually
tested in each of a large number of models.  For datasets that 
result in many unstable models,
this procedure may result in a time-consuming and
perhaps inefficient computational effort.  Here, we follow the 
methodology of \cite{verfor2009},
which relies on MCMC-based simulations to derive
ensembles of initial conditions.  

We assume that the motion is described by a sum of Keplerian 
(i.e. non-interacting) orbits.  Each ensemble of initial conditions 
consists of a list of masses, semimajor axes, eccentricities,
arguments of pericenter, inclinations and nodes, assuming an 
initial mean anomaly for each planet.  We then assign 
line-of-sight and relative inclinations to planets
with these ensembles of orbital elements in order
to sample from the entire phase space of possible initial conditions.

When planetary orbits are integrated forward in time, the system may 
exhibit a variety of evolutionary
paths. Unrestricted planetary inclinations admit a wide variety of phase
space regimes, often including those where apsidal and resonant angles 
circulate, librate and exhibit chaotic motions
\citep{micetal2006b}.  
In this work, because we consider systems 
containing two massive exoplanets with a wide range of 
eccentricities and inclinations,
we must appeal to numerical integrations.

Here, we investigate the dynamical evolution of five two-planet systems
which contain sufficiently numerous and accurate radial velocity orbital data
suitable for a wide-ranging study. In \S \ref{results}, we show that the
stability and dynamical properties of these systems can be 
significantly influenced by the initial values of $i_{rel}$ and $i_{LOS}$.  
Four of these systems are ``hierarchical'', meaning they contain large 
ratios of orbital distances, which often include a close-in planet whose evolution
could be affected by the general relativistic precession of the pericenter.
We aim to characterize the eccentricity variation, stability, secular effects
and resonant signatures of planets in these systems as a function of 
$i_{rel}$ and $i_{LOS}$, while taking into account the uncertainties 
of the measured orbital parameters for each planet.  
We combine investigations of the hierarchical HD 12661 with the system
studied here to draw inferences
for effective methods of future dynamical investigations of hierarchical systems,
and to determine their general evolutionary trends (such as the propensity, 
or lack thereof, for planets to periodically attain circular orbits; 
\citealt*{bargre2006a,bargre2006b}).

We describe our methodology in \S \ref{methods}, and present
the results of the N-body simulations in \S \ref{results}.
Tables \ref{Tabe11964}-\ref{Tablib190360} present
summary statistics, two or three tables per system, arranged by row
according to the binned relative inclination between both planets;
the accompanying figures help the reader visualize the data.
We discuss the results in \S \ref{discussion} and
conclude in \S \ref{conclusion}.

\section{Methods} \label{methods}

The setup of our simulations closely follows that 
of \cite{verfor2009}.  First, we generate a sample
of initial conditions that will serve as the basis
for numerical integrations.

Using RV data from 
\cite{wrietal2008}, we utilize the MCMC
analysis methods and prior distributions 
from \cite{ford2006}.  In particular,
we calculate 5 Markov chains each
containing $10^5$ or $10^6$
states per system.  Each state includes 
the orbital period ($P$), velocity amplitude
($K$), eccentricity ($e$), argument of pericenter measured from the
plane of the sky ($\omega$), and mean anomaly at a given epoch ($u$)
for each planet specified by subscripts.  We randomly select 
13 sets of 500 states,
where each set is restricted to a particular range
of $i_{rel}$.  
We sampled from a uniform distribution
for the longitude of ascending nodes, $\Omega$, and from 
an isotropic distribution for the $i_{LOS}$ of each planet. 
Then we used rejection sampling to choose
500 isotropically distributed pairs of $i_{LOS}$
values in each of 13 $i_{rel}$ bins.
The planet masses, $m$,
and semimajor axes, $a$, are obtained from each set of
($P,K,e,\omega,i,\Omega,u$) values from relations
derived with a Jacobi coordinate system
\citep{leepea2003}.
The 13 $i_{rel}$ bins are split into
one coplanar bin and 12 bins which contain systems 
with relative inclinations in $15^{\circ}$ intervals
from $0^{\circ}$ to $180^{\circ}$. We used stratified sampling 
to obtain sufficiently large samples at both low
and high inclinations.

Second, we perform integrations to test each set of 
initial conditions for long-term stability.  We integrated 
each set of initial conditions with
the hybrid symplectic integrator of {\tt Mercury}
\citep{chambers1999} for 1 Myr, and obtained snapshots
of the system states every $10^3$ yr. Inspection of a 
representative sample of stable configurations
reveals that 1 Myr typically exceeds secular timescales. 

We assume that the actual systems will not
undergo instability immediately ($< 10^6$ yr),
and generate a subsample of initial conditions that 
do not show indications of instability during our
simulations.  We 
classified systems as ``unstable'' if, for either planet,
$(a_{\rm max}-a_{\rm min})/a_0 > \tau a_0$, where $a_{\rm max}$,
$a_{\rm min}$ and $a_0$ represent the maximum, minimum 
and initial values of the semimajor axis,
and $\tau=0.3$.  Additionally we consider systems which have
undergone a close encounter (where two bodies come within a Hill radii
of one another) or collision to be unstable.
Our criterion may not identify
a small fraction of some systems that will manifest instability on longer 
timescales, but importantly it also avoids 
miscategorizing a system exhibiting bounded chaos as unstable
(e.g., \citealt{gozdziewski2003}).

In addition to testing for stability, we compute
the extent to which planets' eccentricities and
mutual relative inclination vary over the course
of the simulations and for what fraction of the planets'
eccentricity and inclination exceed their initial values.
We also determine how closely
the planetary orbits approach circularity by computing 
$\kappa_b \equiv e_{b,min}/e_{b,max}$, where 
$e_{b,min}$ and $e_{b,max}$ represent the minimum and
maximum eccentricities attained throughout a simulation
for planet $b$.  A similar definition holds for planet $c$.
If $\kappa \approx 0$, a planet's eccentricity will
periodically approach zero.  

For two planets on nearly planar orbits, the  
difference of the longitude of periapses, $\Delta\varpi$, has
dynamical significance \citep{chietal2001,chimur2002,beaetal2003,
jietal2003,zhosun2003,namouni2005,miggoz2009a}.
In the planar case, when this difference is $0^{\circ}$
($180^{\circ}$), the planetary orbits are said to be
``aligned'' (``anti-aligned'').  An  ``Apsidal Corotation Resonance''
\citep[ACR;][]{beamic2003,feretal2003,lee2004,kleetal2005,beaetal2006,
micetal2006a,voyhad2006,sanetal2007,micetal2008a,micetal2008b}
refers to system evolutions that result in $\Delta\varpi$ librating
about $0^{\circ}$ or $180^{\circ}$ and can result from
either purely secular evolution or secular plus resonant
interactions. However, for nonzero inclinations, the angle between the 
pericenter directions deviates from $\Delta\varpi$ in the planar case.
The most natural plane on which to measure three-dimensional dynamics
is the invariable plane.  Therefore, for each snapshot in
time and for each N-body integration, we measure  $\Delta\varpi'$, the difference
of the longitude of pericenters after projecting onto the invariable plane.
A time series of these measurements of $\Delta\varpi'$ over 
a period of time, $t_0$, sometimes reveals either ``circulation,''
or ``libration'' about a particular value, the ``libration center,''
$l_0$.  Strictly, an angle librates if there is a range of values
that the angle avoids; $l_0$ is then the middle value of
the range {\it not} avoided by the angle. Unless the angle
is continuously sampled, there will always be a finite range avoided
by the angle.  Therefore, because of finite sampling from integration
output, we consider $\Delta\varpi'$ to librate if every value of 
$\Delta\varpi'$ over a time $t_0$ avoids a range larger
than $90^{\circ}$.  The location of this range depends
on the value of $l_0$ considered.

When testing for libration, we consider a putative value of $l_0$ and
determine if every value of $\Delta\varpi'$ is contained in the range 
from $l_0 - 135^{\circ}$ to $l_0 + 135^{\circ}$.  If an angle librates,
investigators often report a libration amplitude about
$l_0$.  \cite{verfor2009} utilize two alternative variation 
measures, the root mean square deviation and mean absolute 
deviation, which are more robust than the amplitude given finite sampling
and additional short-term perturbations.
We calculate the RMS deviation of $\Delta\varpi'$ about
each of $l_0 = (0^{\circ}, 90^{\circ}, 180^{\circ}, 270^{\circ})$.
Our choice of $l_0$ includes each of the common libration centers.
Analytical considerations \citep{murder1999,morbidelli2002} 
demonstrate that $l_0 = 0^{\circ}$ and $\l_0 = 180^{\circ}$ 
(``symmetric solutions'') are the
only possiblities for purely secular evolution, and numerous numerical studies of
aligned and anti-aligned configurations bear this trend out.
However, $l_0$ may take on other values 
\citep[``asymmetric solutions'';][]{beauge1994,
winmur1997,janetal2002,beaetal2003,murchi2005} if a system
is near a mean motion resonance.  Of the systems considered here,
only HD 108874 is near a MMR.  Assuming the system is not
undergoing asymmetric libration and considering a few alternative
values of $l_0$ (e.g., $90^{\circ}$ and $270^{\circ}$) allows us to estimate the frequency
of systems being misclassified.

\section{Results}  \label{results}

\subsection{Common Attributes}

\cite{wrietal2008} provide  
best-fit orbital solutions to radial velocity
observations of multiplanet systems.
Our calculations are based on the posterior
probability distribution for orbital parameters
described in Section \ref{methods}.  Here, we
give an approximate summary of system properties.
In all systems except HD 190360,
planet $b$ is the inner planet.  The ratio of semimajor
axes nearly exceeds one order of magnitude in 
HD 11964 ($a_b,a_c \approx 0.2, 3.2$ AU),
HD 38529 ($a_b,a_c \approx 0.1, 3.7$ AU),
HD 168443 ($a_b,a_c \approx 0.3, 2.9$ AU), and
HD 190360 ($a_c,a_b \approx 0.1, 3.9$ AU),
and approximately corresponds to the $4$:$1$
Mean Motion Resonance (MMR) in HD 108874 
($a_b,a_c \approx 1.1, 2.7$ AU).  Assuming coplanarity,
The outer planet is more massive in 
HD 11964 ($M_b \sin{i_{b,LOS}},M_c \sin{i_{c,LOS}} \approx 0.1,0.7 M_{{\rm Jup}}$),
HD 38529 ($M_b \sin{i_{b,LOS}},M_c \sin{i_{c,LOS}} \approx 0.8, 13 M_{{\rm Jup}}$), 
HD 168443 ($M_b \sin{i_{b,LOS}},M_c \sin{i_{c,LOS}} \approx 8, 18 M_{{\rm Jup}}$), and
HD 190360 ($M_c \sin{i_{c,LOS}},M_b \sin{i_{b,LOS}} \approx 0.06, 1.5 M_{{\rm Jup}}$), whereas
the planet masses are comparable
in HD 108874 ($M_b \sin{i_{b,LOS}},M_c \sin{i_{c,LOS}} \approx 1.4, 1.0 M_{{\rm Jup}}$).
In the limit that a planet's orbit is seen face-on, the planet's
mass diverges and the corresponding system becomes unstable.
However, most orbital orientations with respect to the Earth
yield actual planet masses less than twice the observed mass; 
a planet's true mass exceeds its observed mass by at least
one order of magnitude only if the planet's orbit is within
$\approx 5.8^{\circ}$ of being face-on.
The currently observed orbital eccentricities of all the planets
studied here vary from nearly circular ($\approx 0.01$) to 
significantly eccentric ($\approx 0.53$).

\cite{libhen2007} analyzed the proximity of particular multi-planet
systems to strong mean motion resonance and found that HD 38529, 
HD 168443, HD 190360 are dominated by non-resonant motions.  For HD 11964,
the semimajor axis ratio of both planets typically exceeds
10, so a mean motion resonance is unlikely to
dominate the evolution of that system.  Of the 5 systems studied, 
only HD 108874 is near a mean 
motion resonance of order $\lesssim 10$.

\subsection{Non-Keplerian Effects}

Because some of the above planets harbor semimajor axes of just
a few tenths of an AU, we must consider the possible effects
on the dynamics due to the pericenter precession caused by 
GR \citep{foretal2000,adalau2006a,adalau2006b}.
This precession is the most significant deviation from a 
Keplerian orbit at these orbital periods and we use
the standard approximation \citep{fabtre2007}:

\begin{equation}
\dot{\omega}_{{\rm GR}} =
\frac{3 G^{3/2} \left( M_{\star} + M_{{\rm inner}} \right)^{3/2}}
{a_{{\rm inner}}^{5/2} c^2 \left( 1 - e_{{\rm inner}}^2 \right)}
\approx 
\frac{3 G M_{\star}}{a_{{\rm inner}} c^2 
\left( 1 - e_{{\rm inner}}^2 \right)} n_{{\rm inner}}
,
\label{GRpre}
\end{equation}

\noindent{where} $n$ refers to the mean orbital motion, $c$ the speed of light,
$M_{\star}$ the mass of the star, and where the subscripts ``inner'' and ``outer'' 
will henceforth refer to the inner and outer planets.
In order to assess the
possible effect of GR on our simulations, we can compare
the value of $\dot{\omega}_{{\rm GR}}$ with the magnitude of the
precession caused by secular evolution with the outer planet \citep{zhosun2003}:

\begin{equation}
\dot{\omega}_{{\rm sec}} = \sqrt{ 
(y_1 - y_2)^2 + 4 y_1 y_2 \left(\frac{b_{3/2}^{(2)}}{b_{3/2}^{(1)}}\right)^2}
\label{secordLL}
\end{equation}

\noindent{where}

\begin{mathletters}
\begin{eqnarray}
y_1 &=& \frac{1}{4} 
         \frac{M_{c} a_{b}^{-3/2}}{\sqrt{M_{\star} + M_{b}}} \alpha^2 
          b_{3/2}^{(1)}\left(\alpha \right)
\label{acoef1}
\\
y_2 &=& \frac{1}{4} 
         \frac{M_{b} a_{c}^{-3/2}}{\sqrt{M_{\star} + M_{c}}} \alpha
          b_{3/2}^{(1)}\left(\alpha \right)
\label{acoef2}
\end{eqnarray}
\end{mathletters}

\noindent{and} $b_{s}^{(j)}\left(\alpha \right)$ are Laplace coefficients, 
with $\alpha$ representing the semimajor axis ratio such that $\alpha < 1$.
Equation (\ref{secordLL}) is derived based on Laplace-Lagrange
secular theory, accurate to second-order in the eccentricities.
The planetary eccentricities in our simulations may attain values close
to unity, and higher-order expansions do not admit eigenvalue
solutions such as those from 
Eq. (\ref{secordLL}) \citep[see, e.g.,][]{verarm2007}.  Further,
by definition, secular theory assumes constant semimajor axes.
Hence, the estimate of $\dot{\omega}_{{\rm sec}}$ for 
potentially near-resonant systems such as HD 108874 may be inaccurate.

With the above caveats in mind, the ratio of these two precession rates, 
$\chi_{{\rm sec}} \equiv 
\dot{\omega}_{{\rm sec}} / \dot{\omega}_{{\rm GR}}$,
provides a measure of the importance of GR in a system.  
Table \ref{Tabchi} lists values of $\chi_{{\rm sec}}$
for representative values of orbital parameters and masses
of all systems studied.  Because $\chi_{{\rm sec}} \gg 1$ only for
HD 168443 and HD 108874, general relativity may play
a significant role in the dynamics in HD 11964, HD 38529 
and HD 190360.  Therefore, we explicitly add into Mercury the 
effect of GR according to the approximation in Eq. (\ref{GRpre}) 
for calculating an additional set of 6500 
simulations for each of the three systems.

For highly inclined systems, a planet may undergo coupled inclination
and eccentricity oscillations through a secular exchange
of angular momentum, known as Kozai oscillations.
\cite{kisetal1998} gives the period for 
these oscillations -- the ``Kozai period'' -- as:

\begin{equation}
P_{{\rm Koz}} = 
\frac{2 P_{{\rm outer}}^2}{3 \pi P_{{\rm inner}}}
\frac{M_{\star} + M_{{\rm inner}} + M_{{\rm outer}}}{M_{{\rm outer}}}
\left(
1 - e_{{\rm outer}}^2
\right)^{3/2}
.
\end{equation}

\noindent{Because} this period refers to the oscillations of
the inner planet's argument of pericenter, the relative 
contribution of the Kozai mechanism to GR can be expressed
as: $\chi_{{\rm Kozai}} \equiv 
\dot{\omega}_{{\rm Kozai}} / \dot{\omega}_{{\rm GR}}
= 2 \pi / P_{{\rm Koz}} \dot{\omega}_{{\rm GR}}$.
Table \ref{Tabchi} reports representative values of
$\chi_{{\rm Kozai}}$ for each system studied here.
The values demonstrate that the Kozai mechanism could be
significant (if the relative inclination exceeds $\approx 40^{\circ}$) 
for HD 168443 and HD 108874, the two systems
where secular perturbations from the outer planet dominates the effect
of GR.

Other possibly significant effects for Hot Jupiters
are tidal deformation and stellar 
oblateness.  In order to assess the magnitude 
of these additional effects, we derive ratios from the 
expressions given by 
\cite{jorbak2008}:

\begin{eqnarray}
\chi_{{\rm tide}} && \equiv  
\frac{\dot{\omega}_{{\rm tide}}}{\dot{\omega}_{{\rm GR}}}
\approx
0.51  
\left[ \frac{1 + (3/2)e_{\rm inner}^2 + (1/8)e_{\rm inner}^4}
{\left(1 - e_{\rm inner}^2\right)^4}\right]
\nonumber
\\
&& \times \left( \frac{M_{{\rm Jup}}}{M_{{\rm inner}}} \right)
\left( \frac{0.05 {\rm AU}}{a_{{\rm inner}}} \right)^4
\left( \frac{R_{{\rm inner}}}{R_{{\rm Jup}}} \right)^5
,
\end{eqnarray}

\begin{eqnarray}
\chi_{{\rm obl}} && \equiv 
\frac{\dot{\omega}_{{\rm obl}}}{\dot{\omega}_{{\rm GR}}}
\approx
0.021 
\left(1 - e_{{\rm in}}^2\right)
\nonumber
\\
&& \times \left( \frac{0.05 {\rm AU}}{a_{{\rm in}}} \right)
\left( \frac{J_2}{10^{-6}} \right)
\left( \frac{M_{\odot}}{M_{\star}} \right)
\left( \frac{R_{\star}}{R_{\odot}} \right)^2
,
\end{eqnarray}

\noindent{}where $R_{\star}$ refers to the 
radius of the star and $J_2$ is a oblateness constant present
in the typical expression for gravitational potential 
(e.g., Eq. 6.218 of \citealt{murder1999}).  Both ratios 
$\chi_{{\rm tide}}$ and $\chi_{{\rm obl}}$ 
are only weakly dependent on the eccentricity
of the inner planet. The ratios
can be estimated only in a rough sense because of the generally unknown stellar
$J_2$ for the former and because of the unknown radii of several planets 
and the strong dependence on $R_{\rm inner}$ for the latter.  Further,
for $\chi_{{\rm tide}}$, not shown are the dependencies on the internal
structures of both the star and planet, which are unknown.
Table \ref{Tabchi} lists estimates for the {\it initial} values of these ratios
by adopting $J_2 = 2 \times 10^{-7}$ \citep{pirroz2003} and
assuming $R_{{\rm inner}} \sim R_{{\rm Jup}}$.
In almost all cases, 
$\chi_{{\rm tide}} << 1$ and $\chi_{{\rm obl}} << 1$,
so GR precession will dominate over tides and the effects from 
stellar oblateness. The only possible exception is for HD 190360, with 
$\chi_{{\rm tide}} = 0.21$. Hence, we neglect these latter two effects in our
simulations.

\subsection{System Stability}

\begin{figure}
\epsscale{0.8} 
\plotone{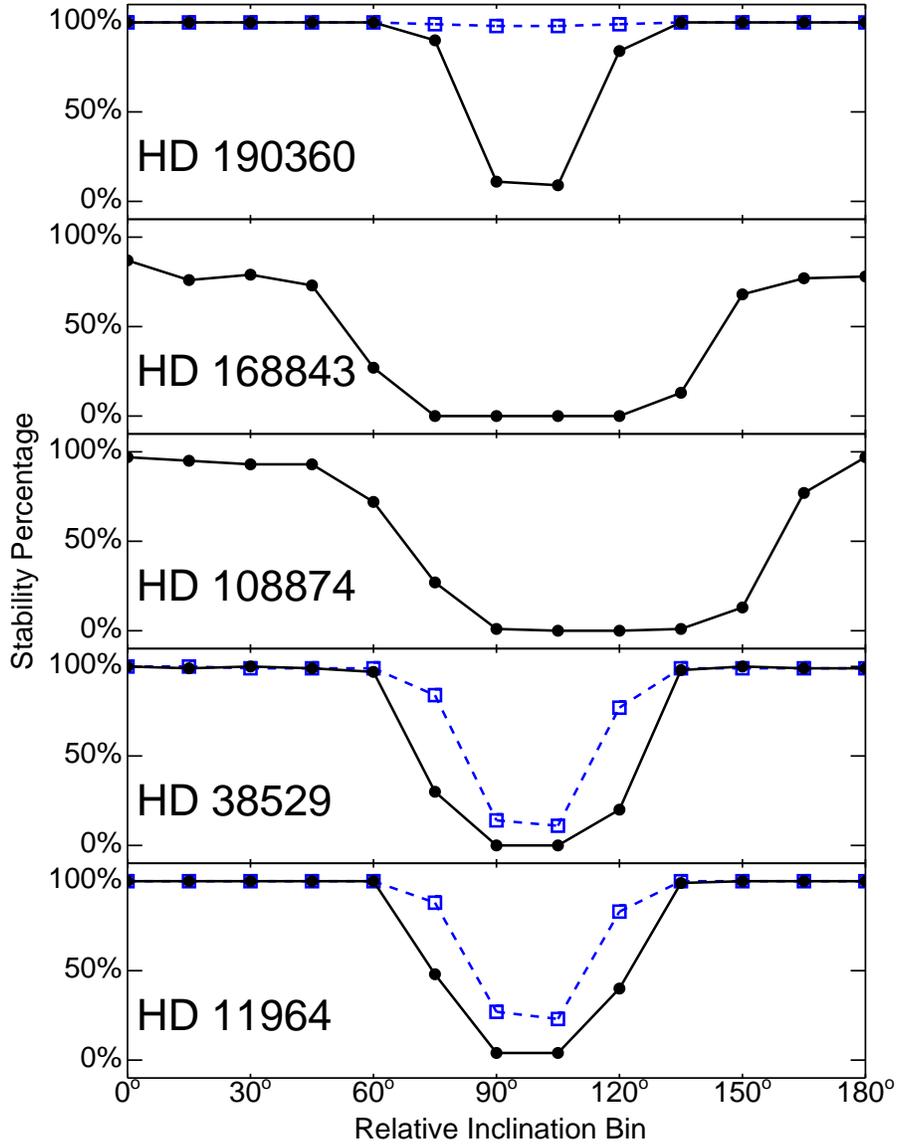}
\figcaption{The percent of stable systems for HD 11964, HD 38529, HD 108874, HD 168443 and HD 190360 as a function of relative inclination when the effect of general relativity is included (blue/dashed lines and squares) and when it is not (black/solid lines and dots).  The data points correspond to $15^{\circ}$-wide relative inclination bins along with a coplanar prograde bin.
\label{stabpanels}}
\end{figure}

For all systems except HD 168443, over $95\%$ of the nearly
coplanar systems, in both the prograde and retrograde sense,
remained stable over $1$ Myr.  Largely independent
of each planet's line-of-sight inclination,
all systems demonstrate that as the relative inclination between
both planets approaches $90^{\circ}$, the systems become 
increasingly likely to undergo instability.  Figure \ref{stabpanels}
graphically depicts the fraction of systems which remain stable
over 1 Myr for all five systems.
This figure broadly quantifies the effect that GR has on
stabilizing systems with a close ($a \lesssim 0.1$ AU) planet.

When Kozai oscillations are present, the maximum eccentricity achieved by
the inner planet ($e_{{\rm inner,max}}$) is lower in our 
numerical simulations which include GR than in the simulations which
do not include GR.  The Kozai 
mechanism induces angular momentum transfer which causes the inner 
planet's eccentricity, inclination and argument of pericenter to 
evolve.  Because these effects are coupled, when GR causes a 
change in the rate of argument of pericenter, the effect affects 
the planet's eccentricity as well.  Both \cite{linetal2000} and 
\cite{fabtre2007} relate the Kozai timescale, 
general relativistic precession, and $e_{{\rm inner,max}}$ through:

\begin{eqnarray}
\cos^{2}i_{{\rm rel,0}}
&=& \frac{3}{5} 
\left(
1 - e_{{\rm inner,max}}^2
\right)
-
\nonumber
\\
&& \frac{2}{5}
\left[
\left(
1 - e_{{\rm inner,max}}^2
\right)^{-1/2}
+ 1
\right]^{-1}
\left( \frac{2 \pi}{\chi_{{\rm Kozai}}}  \right) \bigg|_{e_{{\rm inner,0}} = 0}
\label{ivseeq}
\end{eqnarray}

\noindent{assuming} that the inner planet is on an initially circular
orbit and where the ``0'' subscript will henceforth refer to an initial value.
We have generalized this equation to include an initial non-zero 
eccentricity (see the Appendix and Eqs. \ref{generale}-\ref{Kozgen}).

Equation (\ref{ivseeq}) demonstrates that for an initial
relative inclination high enough to trigger the Kozai mechanism,
the maximum eccentricity of the inner planet decreases as
$\dot{\omega}_{GR}$ increases.  Hence, a high value of  
$\dot{\omega}_{GR}$ provides a stabilizing mechanism for a system.

\begin{figure}
\epsscale{0.8} 
\plotone{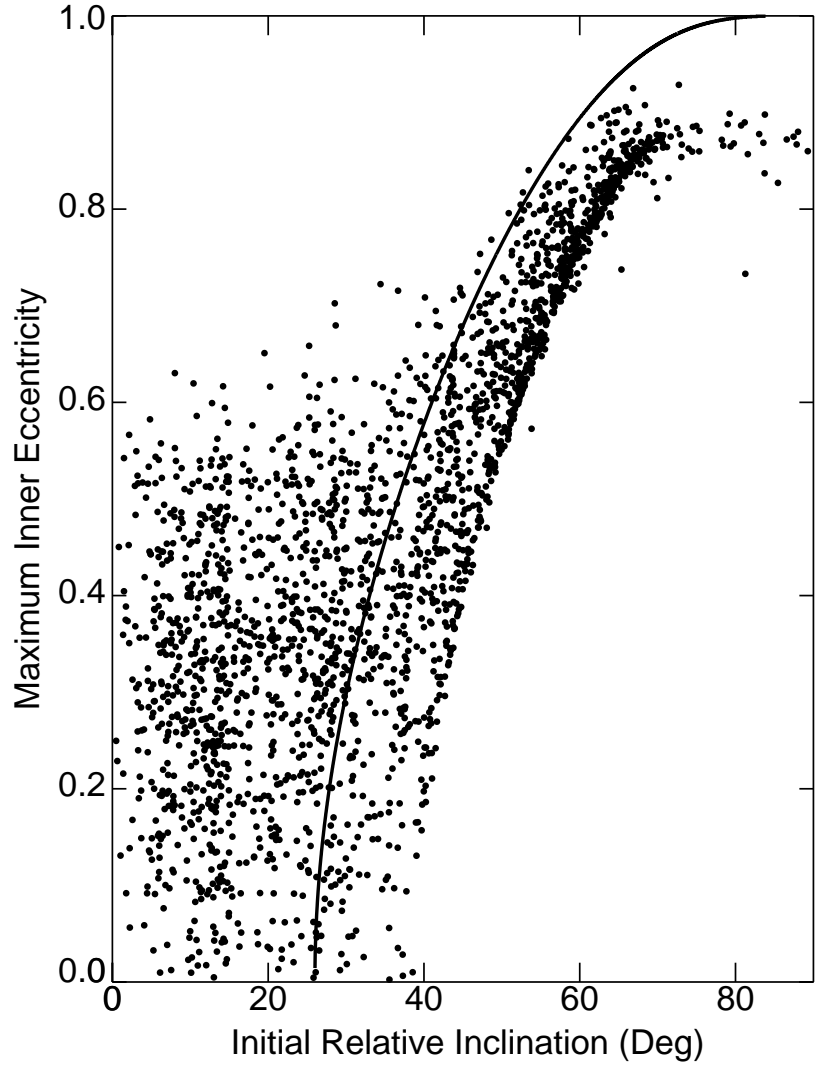}
\figcaption{The maximum eccentricity achieved by the inner planet of
HD 11964 for stable systems as a function of initial relative
inclination (dots).  Overlayed is the solid curve predicted from 
theory (Eq. \ref{generale}) which predicts the maximum 
eccentricity attained by planets affected by general relativity, assuming
an initial representative inner planet eccentricity of 0.28.  Dots appear
above the curve at low inclinations because in this regime, the eccentricity
oscillations are not dominated by Kozai effects, but rather by 
classical Laplace-Lagrange secular theory.
\label{evsi11964}}
\end{figure}

We test the quality of the analytical approximation afforded
by Eq. (\ref{generale}) by using our simulation output. Applying
the values in Table \ref{Tabchi} to Eq. (\ref{Kozgen}) illustrates
that for all systems except HD 190360, the inner planet's 
maximum eccentricity should not be reduced by
GR and Kozai effects. Fig. \ref{evsi11964} plots this eccentricity
difference vs. the initial relative inclination
for simulation data (dots) and Eq. (\ref{generale}) (solid line)
for HD 11964.  The curve in Fig. \ref{evsi11964} is based on a single
representative value of $\chi_{{\rm Kozai}}$; in reality, each dot corresponds
to a different curve.  The dots are generally below
by the curve given in the analytical model, as expected.

The relative contribution of Kozai-like oscillations can be identified by 
the libration of the argument of pericenter of the inner planet 
($\omega_{{\rm inner}}$).
Asymmetric libration of this angle, particularly about the values $90^{\circ}$
and $270^{\circ}$, is characteristic of the Kozai mechanism.  Therefore,
we compute the fraction of inner planet argument of pericenters measured
{\it with respect to} (as opposed to projected on) the invariable plane,
for all systems.  

\begin{figure}
\epsscale{0.8} 
\plotone{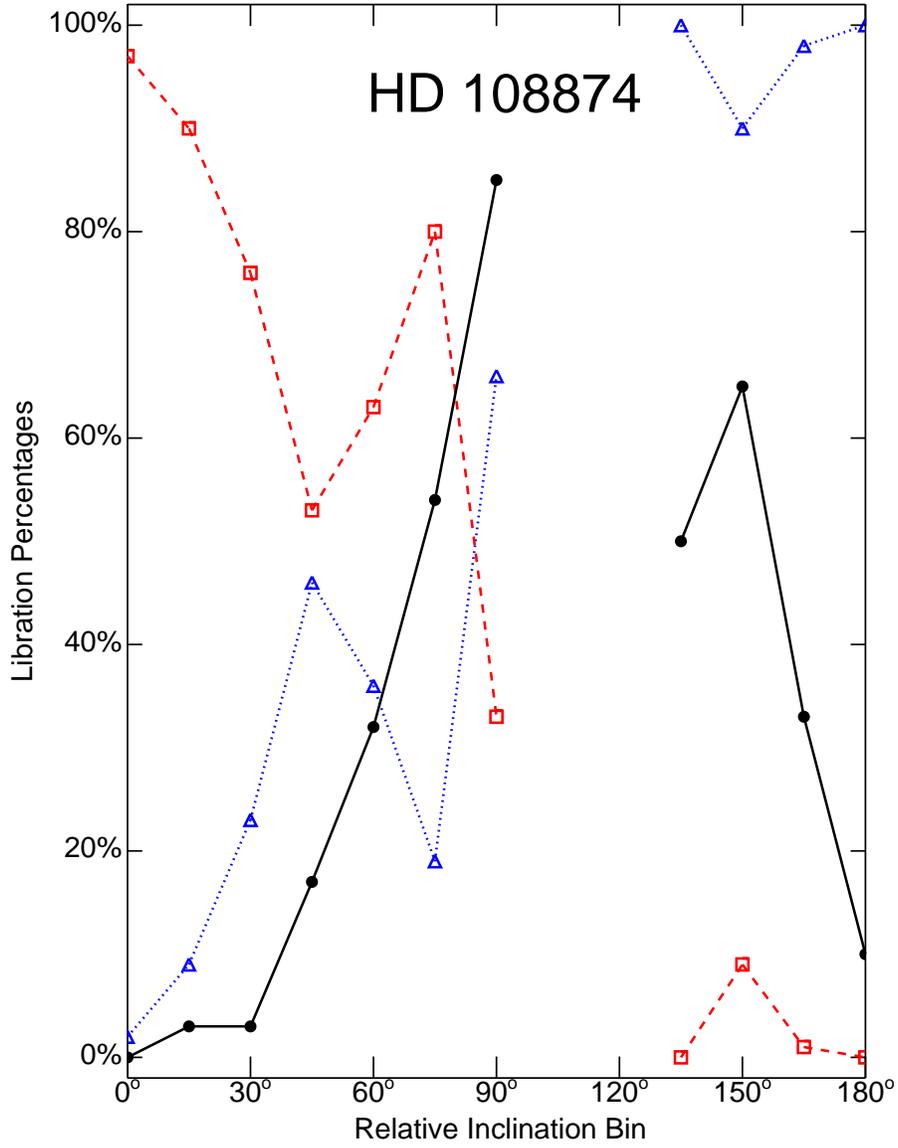}
\figcaption{The percent of simulated HD 108874 stable systems 
whose apsidal angle projected on the invariable plane ($\Delta\varpi'$) is predominantly aligned 
(blue/dotted line with triangles) and antialigned (red/dashed line with squares).  Overlayed
is the percent of systems whose inner planet argument of pericenter measured with
respect to the invariable plane librates around $90^{\circ}$ or $270^{\circ}$ in
a Kozai-like fashion (black/solid line with dots). The data points correspond to $15^{\circ}$-wide relative inclination bins along with a coplanar prograde bin.  Bins with no symbols or lines contain
no stable systems.
\label{libfracs}}
\end{figure}

\begin{figure}
\epsscale{1.0} 
\plotone{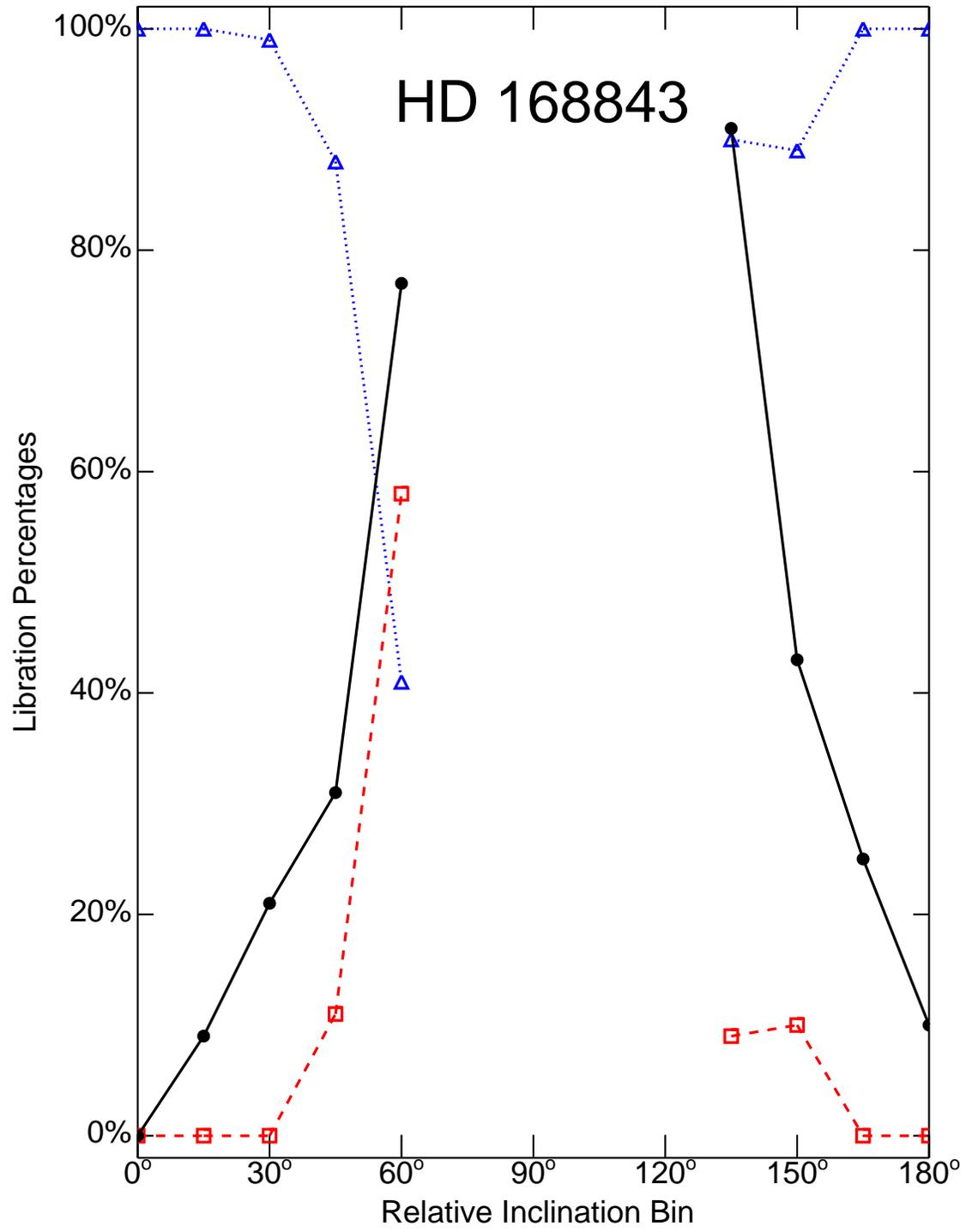}
\figcaption{Same as Fig. \ref{libfracs} but for HD 168843, a system
with two planets not near MMR.
\label{libfracs2}}
\end{figure}

HD 108874 and HD 168443 perhaps best demonstrate the interplay between 
these angles because for those systems (one near-resonant and one far
from strong MMR), $\chi_{{\rm Kozai}} \gg 1$.
Figures \ref{libfracs}-\ref{libfracs2} illustrate the fraction of systems where 
$\Delta\varpi'$ librates about $0^{\circ}$
(blue/dotted lines with triangles) or $180^{\circ}$ (red/dashed line with squares),
and where $\omega_{b}$ librates
around  $90^{\circ}$ or $270^{\circ}$ (black/solid line with dots).  Prograde and
retrograde coplanar systems predominately demonstrate antialignment 
and alignment of $\Delta\varpi'$, respectively, for HD 108874, and alignment
only in HD 168843.   The solid lines peak as
$i_{{\rm rel}}$ approaches $90^{\circ}$.  Therefore, the Kozai mechanism has a significant
effect at large relative inclination values, as demonstrated by the asymmetric 
libration of $\omega_{b}$.  This libration
would be masked by considering libration of $\Delta\varpi'$ alone, which would 
instead suggest that about half of the systems show alignment and the other
half show antialignment at high relative inclinations.

\subsection{HD 11964}  

\cite{butetal2006} announced the existence of the multiple planet
system around HD 11964.  Soon after, \cite{gregory2007b} performed
a Bayesian analysis of the same data and presented evidence 
for a third planet (of $\sim 0.21 M_J$) in the system at $\sim 1.1$ AU.
This planet's existence would imply an unusually small jitter of less than
$0.9$ m/s.  Further, it would require that the commensurability with a one year period 
is not the result of aliases with the annual observing pattern and/or
imprecise barycentric correction \citep{baluev2008}.  Therefore, 
there is only strong evidence for two planets 
given the current data, and we consider the dynamics 
of two-planet solutions.

We ran 6500 N-body integrations of HD 11964 including GR
and 6500 integrations not including GR.  We present 
the simulation output in Tables \ref{Tabe11964GR}
and \ref{Tablib11964}.  Table \ref{Tabe11964GR} indicates:
1) The median amplitude of the outer planet's current eccentricity 
does not exceed $0.075$ for any inclination.  The value of $\kappa_c$
($\equiv e_{c,min}/e_{c,max}$),
indicates that the eccentricity variation achieved throughout 
the simulation is modest; $\kappa_c \ge 0.83$ in all cases with GR.
Regardless of the system's initial orientation, the planet's orbit 
does not become significantly more circular during its secular evolution.   
2) In the coplanar prograde and retrograde (general relativistic) cases, 
the eccentricity evolution is quite modest 
(e.g. $\kappa_b \approx 0.92 (0.95), 0.89 (0.93)$ and
$\kappa_c \approx 0.92 (0.94), 0.93 (0.93)$).
3) We find no indication of instability
except when $60^{\circ} < i_{{\rm rel,0}} < 120^{\circ}.$  
When $i_{{\rm rel}}$ approaches $90^{\circ}$, the eccentricity 
of the inner planet begins to vary drastically.
Incorporating GR precession decreases the inner planet's 
eccentricity variation enough to nearly double the 
fraction of stable systems when 
$60^{\circ} < i_{{\rm rel,0} } < 75^{\circ}$ and $105^{\circ} < 
i_{{\rm rel,0}} < 120^{\circ}$, and quintuple the fraction
of stable systems when $75^{\circ} < i_{{\rm rel,0}} < 105^{\circ}$.

Table \ref{Tablib11964} indicates:  4) In the coplanar prograde and
retrograde cases, $\Delta\varpi' > 90^{\circ}$, indicative of 
large amplitude libration (regardless of the inclusion of GR).
5) About three-quarters
of systems with $45^{\circ} < i_{{\rm rel,0}} < 75^{\circ}$ and 
$105^{\circ} < i_{{\rm rel,0}} < 135^{\circ}$ demonstrate 
asymmetric libration of $\omega_b$, indicating that the 
Kozai mechanism does play a significant role at these 
high relative inclinations.
6) Secular perturbations lead to complex dynamics at both low and high inclinations.
For moderate relative
inclinations ($30^{\circ} < i_{{\rm rel,0}} < 45^{\circ}$, 
$120^{\circ} < i_{{\rm rel,0}} < 135^{\circ}$), $\Delta\varpi' \approx 64^{\circ}$
and is just as likely to librate about $0^{\circ}, 180^{\circ}$, or some
other value.  These relative inclinations
produce large RMS amplitudes, and in the prograde case,
demonstrate preferential alignment of $\Delta\varpi'$.

\subsection{HD 38529}

The planet HD 38529 b was announced in \cite{fisetal2001} and
the outer planet (HD 38529 c) was confirmed in \cite{fisetal2003}.  
HD 38529 c has a minimum mass within tenths of a Jupiter mass of
the oft-cited 13 $M_{{\rm Jup}}$ cutoff. In addition to
orbiting planets, the host star exhibits
IR excess, indicating the presence of a cool disk at wide 
separations \citep{moretal2007}.  

Tables \ref{Tabe38529GR} and \ref{Tablib38529}
summarize the results of all 13,000 simulations for HD 38529.
The first two tables demonstrate: 1) For relative inclinations
within $60^{\circ}$ of coplanarity, the vast majority of systems
are stable.  2) Almost all systems that are within $15^{\circ}$
of $i_{{\rm rel,0}} = 90^{\circ}$ become unstable when neglecting GR;
including GR increases the number of stable systems by over
a factor of 50 to $11\%-14\%$.  3) The outer planet's eccentricity remains
relatively constant (to within $10\%$) in all GR-based cases (and
all non-GR-base cases except for $i_{rel,0}$ values closest to
$90^{\circ}$).  Recall that GR primarily contributes to the pericenter
change of the {\it inner} planet.  Hence, the effect on the outer
planet is indirect, yet non-negligible at high relative inclinations.  
4) The eccentricity variation of 
the inner planet increases gradually as $i_{{\rm rel,0}}$ approaches $90^{\circ}$.  
Including GR slightly decreases the variability of the inner planet's
eccentricity. 5) For highly inclined, stable systems, the relative inclination 
can vary from less than $90^{\circ}$ (prograde) to greater than
$90^{\circ}$ (retrograde).

Regarding libration properties, Table \ref{Tablib38529} 
illustrates:  6) The RMS deviations of $\Delta\varpi'$ are 
high enough to be consistent with
circulation in all relative inclination bins, 7) Assuming 
libration of $\Delta\varpi'$,
alignment is strongly preferred in the
coplanar prograde and retrograde cases. 8) GR's
primary effect on the libration of $\Delta\varpi'$ is to cause
preferential alignment of the angle at moderate relative 
inclinations 
($45^{\circ} \le i_{{\rm rel,0}} \le 75^{\circ}$ and 
$105^{\circ} \le i_{{\rm rel,0}} \le 135^{\circ}$).  9) For these
values of $i_{{\rm rel,0}}$, asymmetric libration of $\Delta\varpi'$ 
is dominant, indicating the role of the Kozai mechanism.

\subsection{HD 108874}

HD 108874's two known planets \citep{vogetal2002,vogetal2005} have a
orbital period ratio close to four, suggesting that they may be in or near a
4:1 MMR \citep{gozdziewski2006,libhen2007}.  Our joint statistical and
dynamical analysis is well suited to testing this possibility.  First,
we performed a standard MCMC analysis assuming that each planet
traveled on an independent Keplerian orbit. As for the previous
planetary systems, we used the standard MCMC output to generate
ensembles of initial conditions for N-body integrations to test for
dynamical stability and secular evolution of the eccentricity and
inclinations.  In principle, these results might be affected by the
planet-planet interactions, particularly if the two planets are in a
mean-motion resonance.  Therefore, we performed a second analysis
using full N-body integrations for both the statistical and dynamical
analyses.  The prior probability distributions were similar to those in our
previous analysis with non-interacting planets.  In other words, the priors are
based on the period and velocity amplitude inferred from the Jacobi
orbital elements.  We adopt a stellar mass of 0.95$M_\odot$.  Since
the relative orientation of the orbits matters for the N-body
simulations, we must add priors for the inclination of the orbital
plane and longitude of nodes (where the orbit punctures the sky
plane). We consider both: 1) a prograde coplanar configuration with an
isotropic prior for the inclination of the system (i.e., a common
inclination and a common longitude of ascending node for each system),
or 2) an isotropic prior for the inclination of each orbit and a
uniform prior for the line of nodes.  As before, we select subsets of
the output based on the relative inclination to be used for long-term
N-body simulations to investigate stability and secular eccentricity
and inclination evolution.  The likelihood is evaluated based on
comparing observations and the simulated stellar velocities (in a
barycentric frame) evaluated at the time of each radial velocity
observation.

We again employ Markov chain Monte Carlo (MCMC) to sample from the
posterior probability distribution for the masses and orbital
parameters.  The convergence rate of MCMC is often sensitive to the
algorithm for proposing the next state (e.g., candidate transition
probability distribution function).  When performing Bayesian analysis
with N-body interactions, we replace the Gaussian random walk
Metropolis algorithm adapted for analyzing radial velocity data sets
(Ford 2006) with a differential evolution Markov chain Monte Carlo
algorithm (DEMCMC; ter Braak 2006).  In this algorithm, each state of the
Markov chain consists of an ensemble of initial conditions (known as a
generation).  For our analysis of HD 108874, each generation consisted of an ensemble
of 1,024 sets of initial conditions.  For the initial generation, the initial conditions were drawn from a posterior sample calculated assuming
independent Keplerian orbits (i.e., using the standard MCMC approach;
Ford 2006).  For each of $10^5$ subsequent generations, 1,024 new sets
of initial conditions are proposed, based on combinations of multiple
sets of initial conditions from the previous generation.  The
acceptance probability is chosen so as to converge to the desired
posterior probability distribution.  As before, we perform long-term
N-body integrations for a subset of the initial conditions drawn from the
latter portions of the Markov chain.

The results of the DEMCMC+N-body analysis (including planet
interactions) and the standard MCMC analysis (assuming independent
Keplerian orbits) are extremely similar (Table \ref{Tabe108874}
For example, for coplanar, prograde systems, the MCMC (DEMCMC+N-body)
analysis estimates the fraction of stable systems to be $97.6\%$
($99.0\%$).  Similarly, the statistics describing the extend of
eccentricity evolution are $\kappa_b = 0.29$ ($0.29$) and $\kappa_c =
0.25$ ($0.26$).  Thus, we conclude that our results for HD 108874 are
not sensitive to the effects of short-term planet-planet interactions
or the assumption of independent Keplerian orbits.  The exception is
when $\sin i$ becomes small.  However, if $\sin i$ is small enough to
cause significant short-term interactions, then the system is very
likely to be rejected based the test for orbital stability.


In order to explore the possibility of the system being in resonance, 
we perform a systematic
search for the possible libration of all resonant angles
associated with the $4$:$1$, $7$:$2$, $11$:$3$, and $15$:$4$
MMRs.  These commensurabilities were chosen based on their proximity
to an orbital period ratio of four.  Each resonant angle has the form

\begin{equation}
\phi = j_1 \lambda_{{\rm out}} +j_2 \lambda_{{\rm in}}
      +j_3 \varpi_{{\rm out}}  +j_4 \varpi_{{\rm in}}
      +j_5 \Omega_{{\rm out}}  +j_6 \Omega_{{\rm in}},
\end{equation}

\noindent{where} $\lambda$, 
represents
the mean longitude, $\varpi$ represents
the longitude of pericenter, and $j_i$ are constants.
We sample four MMRs, up to order 11, in case any 
high- $(> 10)$ order resonances may play a role in 
the system's dynamical evolution \citep{micetal2006b}. 
For each of the four MMRs, 
all combinations 
of $j_i$ are considered such that $\sum j_i = 0$ and 
$(j_5 + j_6)$ is even.  We performed this search for 
all 6,500 systems simulated for HD 108874.  We flagged
any angles which librated for at least $t_0 = 10^5$ yr.
We discovered that less than $0.5\%$ of all 
simulations studied exhibited
libration, and even then only for a few $10^5$ yr.  There is 
no apparent correlation with resonance and stability
nor resonance and initial relative inclination.
Thus, HD 108874 is very unlikely to be in resonance based on 
the current RV data.
 
Regardless of the possible influence of resonant or 
near-resonant terms, the characteristics of the secular evolution 
(Table \ref{Tabe108874}) are striking:  
1) Despite a median initial
planet $b$ eccentricity of $\approx 0.12$, $\kappa_b < 0.30$ for
every relative inclination bin, and approaches zero as
$i_{{\rm rel,0}}$ approaches $90^{\circ},$  2) For 
$i_{{\rm rel,0}} = 40^{\circ}-75^{\circ},$ the eccentricity 
of the planetary orbits approaches zero, 3)
Unlike for the inner planet's orbit, the outer planet's orbit demonstrates
an increasingly restrictive eccentricity range
in the retrograde case as the planets' orbits become coplanar.
The values in Table \ref{Tablib108874} reveal
other interesting aspects of the motion:  4) $\Delta\varpi'$
always demonstrates symmetric libration, regardless of $i_{{\rm rel,0}}$.
5) Prograde coplanar systems are almost entirely anti-aligned,
whereas retrograde coplanar systems are entirely aligned.  This
shift from aligned to anti-aligned could indicate a significant
contribution from near-resonant terms even though the
planets are not strictly locked in a MMR.  
6) Prograde coplanar systems feature the smallest libration 
amplitudes. 7) As demonstrated in Fig. \ref{libfracs}, the system
evolution exhibits libration of $\omega_b$ in a robust
Kozai-like fashion for $40^{\circ} < i_{{\rm rel,0}} < 140^{\circ}$
except for the unstable systems prevalent 
in the $90^{\circ} < i_{{\rm rel,0}} < 120^{\circ}$ regime.

\subsection{HD 168443}

After the discovery of HD 168443 b \citep{maretal2001},  
\cite{udretal2002} found a companion brown dwarf
in the system.  \citeauthor{verarm2007}'s (\citeyear{veras2007}) 
numerical integrations
of the system, assuming coplanarity, demonstrate that both
planets undergo significant eccentricity evolution.  They found that the
two most apparent modulation frequencies are $\sim 10^3$ yr and 
$\sim 10^4$ yr, and that the apsidal
angle clearly circulates.

Our 6,500 simulations of this system are summarized
in Tables \ref{Tabe168443} and \ref{Tablib168443}.
Table \ref{Tabe168443} demonstrates: 1) Almost no stable systems
can exist for $60^{\circ} \le i_{{\rm rel,0}} \le 120^{\circ}$ and few
systems remain stable when this range is extended by $15^{\circ}$
in each direction. 2) At least $20\%$ of all systems become unstable
for every relative inclination bin except for the perfectly coplanar
case, where $13.0\%$ of the systems sampled become unstable.  3)
For the vast majority of stable systems, planet $c$'s eccentricity 
variation is independent of the relative inclination.  4) For 
highly inclined, stable systems, the relative inclination 
can vary from less than $90^{\circ}$ (prograde) to greater than
$90^{\circ}$ (retrograde).  Table \ref{Tablib168443} shows 
that: 5) For all relative inclinations, the libration 
amplitude of $\Delta\varpi'$ exceeds $90^{\circ}$, 6) $\Delta\varpi'$  is nearly
always preferentially aligned and 7) $\omega_b$ exhibits strong Kozai-like
behavior for $i_{{\rm rel,0}}$ values just under $60^{\circ}$ and just over
$120^{\circ}$.

\subsection{HD 190360}
\cite{udretal2003} announced a planet in HD 190360 
and \cite{vogetal2005} and \cite{gozmig2006} followed up with
a detection and confirmation of an additional planet
whose mass is approximately one Neptune-mass.  Primarily
because of the inner planet's small semimajor axis, 
GR has the greatest effect on this system
(see Table \ref{Tabchi}). 

Tables \ref{Tabe190360}, \ref{Tabe190360GR} and \ref{Tablib190360}
provide the summary output from the 
system's 6,500 GR-based and 6,500 non-GR-based simulations.
Table \ref{Tabe190360} demonstrates: 1) An appreciable fraction
of stable systems exist for all relative inclinations. 
However, for $60^{\circ} \le i_{{\rm rel,0}} \le 120^{\circ}$,
$\approx 89\%$ of systems without GR are unstable.
Including GR, however, 
stabilizes {\it nearly every one} of those systems.
2) The outer planet's eccentricity varies by less than $1\%$ in the vast majority 
of individual stable systems.  3) The presence of GR reduces the variation
of the inner planet's eccentricity by up to a factor of 4.

Table \ref{Tablib190360} suggests:  
4) Without GR, the lowest libration amplitudes and the highest 
amplitude standard deviations of $\Delta\varpi'$
occur when the relative inclination
is $15^{\circ} - 30^{\circ}$ offset from either the prograde
or retrograde planarity.  
Including GR 
results in different conclusions; the system does not demonstrate 
small libration amplitudes nor asymmetric libration of  $\Delta\varpi'$.
5) In the coplanar and
retrograde planar cases, alignment of $\Delta\varpi'$ is preferred over
anti-alignment by a factor of nearly four.  6) With the presence of GR,
the libration amplitudes all exceed $90^{\circ}$, commensurate
with circulation.  7) The Kozai signatures are weaker than in any other
system, corroborating the trend exhibited by the values of $\chi_{{\rm Kozai}}$ 
from Table \ref{Tabchi}.

\subsection{HD 12661}

\cite{verfor2009} performed a similar MCMC-based statistical analysis
for HD 12661, but with broader relative inclination bins.  
For completeness, we compare some of their results with the 
outcomes presented here.  HD 12661
is a two planet system where the inner (outer) planet properties are:
$M \approx 2.3 M_{{\rm Jup}} (1.8 M_{{\rm Jup}})$, 
$a \approx 0.8 {\rm AU} (2.4 {\rm AU})$, and
$e \approx 0.35 (0.05)$.  The authors demonstrated that the 
outer planet spends over $\sim 96 \%$ of the time
following an orbit that is more eccentric than the
one currently observed. In the coplanar prograde case, 
all systems were stable, and in the coplanar retrograde case, 
almost all systems were stable.  $16.4\%$
of systems remained stable for 
$60^{\circ} \le i_{{\rm rel,0}} \le 90^{\circ}$,
whereas only $0.2\%$ did so for 
$90^{\circ} \le i_{{\rm rel,0}} \le 120^{\circ}$.  This
asymmetry about $90^{\circ}$ is significant and much
greater than the slight asymmetries about $90^{\circ}$
seen for HD 11964, HD 38529, and HD 190360, but comparable
to those in HD 108874.
Further, the alignment of $\Delta\varpi'$ for HD 12661 
shows a strong dependence on $i_{{\rm rel,0}}$; only $\approx 33\%$ of 
coplanar prograde systems show alignment of $\Delta\varpi'$,
whereas $\approx 99.4\%$ of coplanar retrograde systems
show alignment.  The only system studied here with a similar
trend is HD 108874, whose $\Delta\varpi'$  is nearly entirely
antialigned in the prograde coplanar case, and entirely aligned
in the retrograde coplanar case.

\section{Discussion}  \label{discussion}

Hierarchical systems can be highly eccentric or inclined (or even retrograde).
Such large eccentricities and inclinations imply that numerical
integrations are often necessary to model the dynamics of 
extrasolar systems. Analytical treatments of two-planet evolution abound 
\citep[e.g.][and references therein]{murder1999,morbidelli2002},
and often rely on approximations to the gravitational potential, 
known as ``disturbing functions,''
The Laplace expansion of the disturbing function 
\citep{ellmur2000} relies on a Taylor expansion about zero eccentricities
and inclinations and is limited in eccentricity
by the Sundman criterion \citep{ferrazmello1994,sidnes1994}.
Although effective for small Solar System bodies, this disturbing
function must be used with care with respect to extrasolar
planets \citep{veras2007}. 
\cite{beauge1996} and \cite{beamic2003} developed a high eccentricity
version of a disturbing function in the planar case
which can reliably model the evolution of massive extrasolar planets 
with eccentricities up to $\approx 0.5$.  Because quadrupole and octopole
expansions are in terms of the semimajor axis ratio, their accuracy
is diminished for weakly hierarchical systems.  For noncoplanar systems,
one can use semi-numerical averaging \citep{micmal2004,micetal2006b,miggoz2009b} 
or adaptions of Gauss' method \citep{touetal2009}, which are valid
for high $e$ and $I$.  However, these studies neglect resonant and short-term
perturbations.  We find that these can be significant for some planetary
systems, even if not in MMR (e.g. HD 108874).

We can compare our secular evolution results with those 
from other investigators.  \cite{bargre2006b} performed a dynamical 
study of 2-planet systems which did not incorporate uncertainties 
in the initial conditions and assumed that each system was coplanar 
and edge-on.   With those assumptions, they found that several of 
these systems had at least one eccentricity that periodically obtains 
a small value.  They claimed that an unexpectedly high frequency of 
systems with ``near separatrix motion" implied that this property couldn't 
be due to planet scattering from initially circular orbits.  Instead, we 
find that once we 1) focus our attention on two-planet systems with published 
RV data and significant secular evolution, 2) avoid systems near MMRs, and 
3) account for the uncertainties in the current orbital elements, only a modest 
fraction show significant secular eccentricity evolution with one planet 
returning to a near circular orbit for all viable inclinations.  This fraction
is equal to $2/6500$ for HD 11964, $1/6500$ for HD 38529, $757/6500$ for HD 108874,
$27/6500$ for HD 168443, and $0/6500$ for HD 190360, assuming that the planet
which returns to or stays on a near circular orbit achieves an eccentricity 1) range
of at least $0.1$, and 2) between $0.01$ and $0.05$ at least once 
every $5 \times 10^4$ yrs.  For HD 108874, if we specify that the 
eccentricity range must be at least $0.2$, then the fraction becomes $488/6500$.

\begin{figure}
\epsscale{0.8} 
\plotone{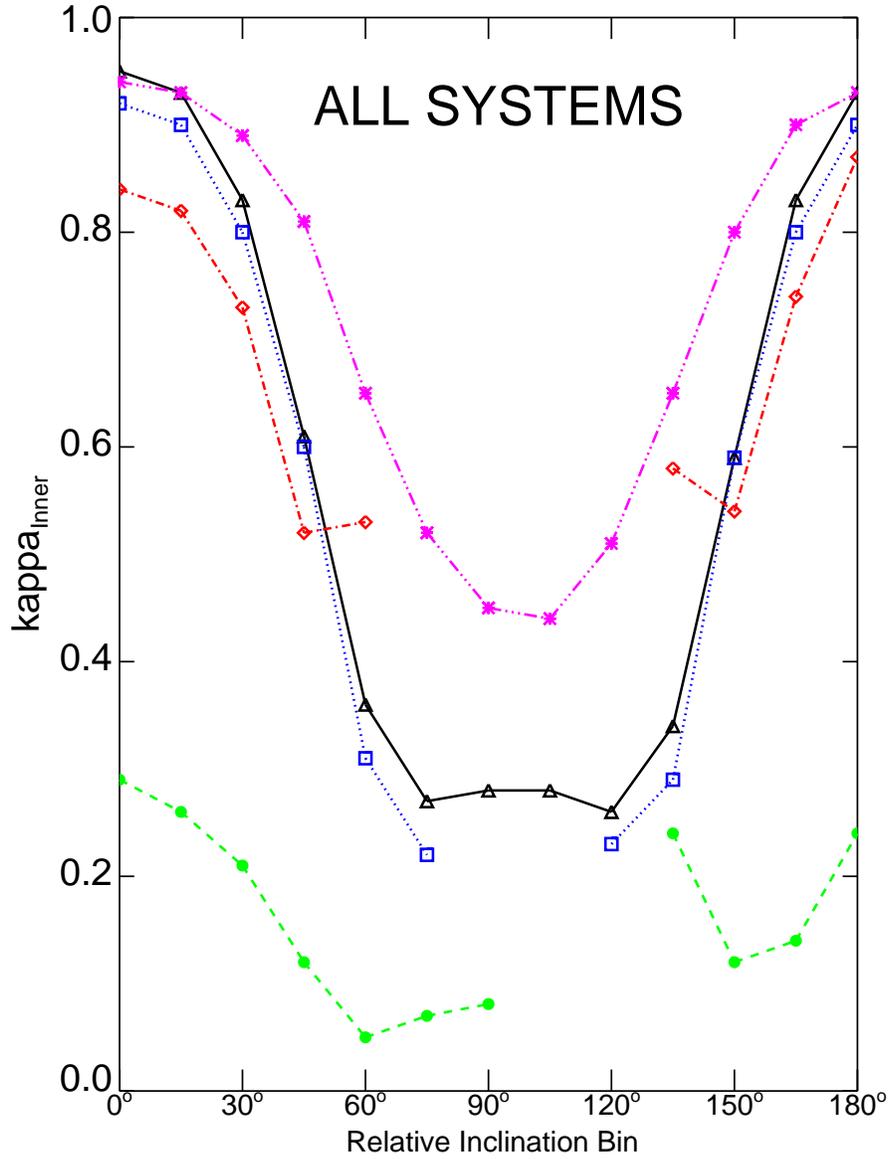}
\figcaption{The averaged eccentricity variation 
($\kappa \equiv e_{min}/e_{max}$) of the inner planet
as a function of initial relative inclination for 
HD 11964 (black/solid lines with triangles), 
HD 38529 (blue/dotted lines with squares), 
HD 108874 (green/dashed lines with dots),
HD 168443 (red/dot-dashed lines with diamonds) and 
HD 190360 (magenta/triple dot-dashed lines with asterisks).
The data points correspond to $15^{\circ}$-wide relative inclination bins along with a coplanar prograde bin.  Bins with no symbols or lines contain
no stable systems.
\label{kapsimple}}
\end{figure}

In particular, Fig. \ref{kapsimple} plots the 
value of $\kappa$ for the five systems analyzed here
as a function of initial $i_{{\rm rel,0}}$.  The figure demonstrates
that the eccentricity variation of the inner planet
typically increases as the initial $i_{{\rm rel,0}}$ approaches $90^{\circ}$,
and that $\kappa$ can take on values from $\sim 0.10-0.90$.
\cite{bargre2006b} use a measure, 
$\epsilon \equiv 2 {\rm min}(\sqrt{u^2 + v^2})/(u_{\rm max} - 
u_{\rm min} + v_{\rm max} - v_{\rm min})$, where
$u \equiv e_{{\rm inner}} e_{{\rm outer}} \sin{(\Delta\varpi)}$
and 
$v \equiv e_{{\rm inner}} e_{{\rm outer}} \cos{(\Delta\varpi)}$
to compare
the minimum eccentricity value of either planet attained during evolution
to the range of variation, assuming the planetary orbits are 
coplanar.  Their value of $\epsilon$ for the near-MMR system
HD 108874 (0.198)
is comparable to the value of $\kappa$ for near-coplanar prograde and retrograde
values (see green/dashed line in Figure \ref{kapsimple}).  However,
their $\epsilon$ values for HD 38529 (0.44), HD 168443 (0.219)
and HD 190360 (0.38) are much less than the values of $\kappa$ for the
near-planar regimes. Thus, we find that hierarchical
systems typically have more moderate eccentricity evolution than suggested
by \cite{bargre2006b}.  Hence, we do not find a significant discrepancy between
the secular evolution of hierarchical two-planet systems and simulations of
planet scattering.   We find that the eccentricities of the inner
planets in these systems vary by a factor of 2 or more for initial 
$i_{{\rm rel,0}}$ values closer to $90^{\circ}$ than to $0^{\circ}$ or $180^{\circ}$.

\cite{libhen2006} have studied the secular evolution of HD 38529
and HD 168443.  The secular eccentricity evolution of the 
outer massive planet in HD 38529 is poorly
modeled by low-order Laplace-Lagrange secular theory \citep{verarm2007};
N-body simulations in the coplanar, edge-on case indicate that 
the inner eccentricity evolves on a timescale of $\sim 10^5$ yr,
and that the apsidal angle circulates.  \cite{libhen2006} estimate
that $\kappa_b = 0.87 (0.90)$ and 
$\kappa_c = 0.999 (0.999)$ for $i_{LOS} = 90^{\circ}$ by using their high-order expansion
(second-order Laplace-Lagrange theory in parenthesis). 
These values compare favorably with
the values in the coplanar row of Table \ref{Tabe38529}, with 
the difference attributed to our isotropic distribution of 
line-of-sight inclinations and 
choice of initial values for the orbital elements.  
We find slightly larger eccentricity evolution but 
qualitatively similar behavior.  \cite{libhen2006} estimate
that for $i_{LOS} = 90^{\circ}$ and using their high-order expansion
(Laplace-Lagrange theory), $\kappa_b = 0.85 (0.89)$, and 
$\kappa_c = 0.83 (0.91)$.  The value of $\kappa_c$ differs significantly 
from our estimate ($0.49$) in Table \ref{Tabe168443}.
We believe this difference is due to the the increased masses 
resulting from our
isotropic line-of-slight inclination distribution,
resulting in a larger extent of eccentricity evolution of the outer planet.

Several researchers have considered the potential for additional small 
planets to exist in these systems, particularly those planets
potentially in the habitable zone.
\cite{jonetal2006} considered
likely habitable zone ranges, $h_z$ and habitability prospects
(yes/no/maybe) for all 6 of the systems studied here.  They found:  
HD 11964 ($h_z = 1.6-3.1$ AU, yes); 
HD 38529 ($h_z = 2.4-4.8$ AU, no);
HD 108874 ($h_z = 1.0-2.0$ AU, no); 
HD 168443 ($h_z = 1.1-2.3$ AU, no); 
HD 12661 ($h_z = 1.0-2.0$ AU, no);
HD 190360 ($h_z = 1.0-2.0$ AU, maybe). 
\cite{erdetal2004} agree that planets in the 
habitable zone of HD 38529 would likely 
have chaotic orbits.  Similarly, both \cite{erdetal2004} and
\cite{barray2004} agree that the prospects for stable
planets in the habitable zone of HD 168443 are slim.
\cite{schetal2007} raise the possibility that dynamically stable Trojan planets 
of HD 108874 b may exist in the habitable zone.

In order to help address the question of whether
HD 190360 may admit habitable terrestrial planets,
we perform 2 additional sets of 1 Myr simulations with a disk of
test particles (a good representation of terrestrial
planets in systems with giant planets).  We
start from the output of MCMC simulations and 
randomly chose 2 sets of stable coplanar initial orbital 
parameters for the system.  For each
system, we superimposed a coplanar disk of 40 terrestrial planets
distributed in semimajor axis according to a Keplerian (power law of $-3/2$)
distribution.  We assigned the eccentricities of the planetesimals 
random values under $0.001$ and randomized their mean anomalies
and longitude of periastrons.  The extent of the planetesimal disk 
was $[1.027, 2.022]$ AU (the same representative habitable zone
values used by \citeauthor{jonetal2006} \citeyear{jonetal2006}) for one
system, and $[0.128 \times (1 + 0.01), 3.92 \times (1 - 0.36)]$ AU 
(the representative range from the apocenter of the inner planet
to the pericenter of the outer planet)  for the other system.
In the first (second) system, we find that 91.4\% (92.7\%) of 
all terrestrial planets survive within the initial planetesimal
disk.  The surviving planets have modest eccentricities: the median
and standard deviation of these values are $0.18 \pm 0.12$ ($0.15 \pm 0.14$).
Most initial instances of instability occur within $10^4$-$10^5$ yr.  
These results indicate that 
that HD 190360 can likely host a stable terrestrial 
planet in the habitable zone.



\section{Conclusion}  \label{conclusion}

Due to limitations of real astronomical observations, often
a significant range of planetary masses is consistent with observations.
Therefore, it is necessary to investigate the dynamics of ensembles of planetary
orbits and masses to accurately model the orbital evolution of exoplanets.
Hierarchical multi-planet systems demonstrate a wide variety of dynamical
behaviors depending on the $i_{LOS}$ and $i_{{\rm rel}}$ values, which are only weakly
constrained by observations.  Inclusion of GR in simulations
of multi-planet systems with a Hot Jupiter may crucially affect the long-term
stability, extent of eccentricity variation, and apsidal configuration directly.  
The eccentricity and inclination evolution of stable highly inclined systems 
are often dominated by Kozai-like oscillations, but can be limited by precession
due to other planets or GR.

\acknowledgments{We thank the referee for helpful comments, Dan Fabrycky 
for assistance in characterizing
the effect of general relativity, and Hal Levison and Jacques Laskar for
useful discussions.  We acknowledge the University of Florida 
High-Performance Computing
Center for providing computational resources and support. 
This research was supported by NASA Origins of Solar Systems 
grant NNX09AB35G and JPL Research Support Agreement \#1326409.
}

\clearpage

\appendix

Because most of our simulations treat planets with non-zero 
eccentricities, we derive a more general relation
than that of Eq. (\ref{ivseeq}).
The two conserved quantities for a two-planet system including GR based
on a quadrupole-order potential are\footnote{Note that Eq. (17) of 
\cite{fabtre2007} should read 
``$H'^2 = \left( 1 - e_{\rm in}^2 \right) \cos^2{i}$.''}:

\begin{equation}
F = 
-\frac{z_1}
{\sqrt{\left(1 - e_{{\rm inner}}^2 \right)}}
+
z_2
\left\lbrace  
-2 - 3  e_{{\rm inner}}^2
+
\left(
3 + 12 e_{{\rm inner}}^2 - 15 e_{{\rm inner}}^2 \cos^2{\omega_{{\rm inner}}}
\right)
\sin^2{I_{{\rm rel,0}}}
\right\rbrace
,
\end{equation}

\begin{equation}
H^2 = z_3 \left(1 - e_{{\rm inner}}^2 \right) \cos^2{I_{{\rm rel,0}}},
\end{equation}

\noindent{where}

\begin{equation}
z_1 = \frac{3 G^2 M_{\star} M_{\rm inner} 
\left( M_{\star} + M_{\rm inner}  \right)}
{a_{{\rm inner}}^2 c^2}
,
\end{equation}

\begin{equation}
z_2 = \frac{G M_{\star} M_{\rm inner} M_{\rm outer}}
{M_{\rm \star} + M_{\rm inner}}
\frac{a_{{\rm inner}}^2}
{8 a_{{\rm outer}}^3  \left( 1 - e_{{\rm outer}}^2 \right)^{3/2} }
,
\end{equation}

\begin{equation}
z_3 = M_{\star} M_{\rm inner}
\sqrt{ \frac{G a_{{\rm inner}}}  
{M_{\star} + M_{\rm inner}} }
.
\end{equation}

Noting that 
$z_1/z_2 = 6 P_{{\rm Koz}} \dot{\omega}_{{\rm sec}} = 12 \pi/\chi_{{\rm Kozai}}$,
one can then compute $F' \equiv F/z_2$ and $H'^2 \equiv H^2/z_3$ from the initial 
conditions and then solve the following implicit equation for $e_{{\rm inner,max}}$:

\begin{equation}
\cos^{2}i_{{\rm rel,0}}
= \frac{3}{5} 
\left(
1 - e_{{\rm inner,max}}^2
\right)
-
\frac{2}{5}
\frac{1 - e_{\rm inner,max}^2}{e_{\rm inner,max}^2 - e_{\rm inner,0}^2}
\left(
\frac{1}{\sqrt{1 - e_{\rm inner,0}^2}} -
\frac{1}{\sqrt{1 - e_{\rm inner,max}^2}}
\right)
\left( \frac{2 \pi}{\chi_{{\rm Kozai}}}  \right)
\label{generale}
\end{equation}

\noindent{Because} Eq. (\ref{generale}) may admit no 
solutions for certain values of $I_{{\rm rel,0}}$,
one must consider the physically plausible solutions.
When the inner planet's initial eccentricity equals zero,
no Kozai librations occur for $\chi_{{\rm Kozai}} < 2\pi/3$.
More generally, as a function of $e_{{\rm inner,0}}$,
no Kozai librations occur when:

\begin{equation}
\chi_{{\rm Kozai}} < 
\frac{4 \pi \left( 1 - \sqrt{1 - e_{{\rm inner,0}}^2 } \right)}
{3 e_{{\rm inner,0}}^2 \sqrt{1 - e_{{\rm inner,0}}^2} }
.
\label{Kozgen}
\end{equation}

\noindent{Equation} (\ref{Kozgen}) demonstrates that 
the critical $\chi_{{\rm Kozai}}$ for which Kozai
oscillations will occur is a weak function of $e_{{\rm inner,0}}$
so that such oscillations should always occur for $\chi_{{\rm Kozai}} > 2 \pi$
when $0 \le e_{{\rm inner,0}} \lesssim 0.79$.  The equations, however,
are just approximations, because they assume  1) secular evolution only (no change
in semimajor axis), 2) a Hamiltonian truncated to the quadruple-order
term, 3) a doubly-averaged Hamiltonian which eliminates short-period
terms, and 4) a fixed outer orbit whose orbital plane does not vary.

\clearpage

\begin{deluxetable}{ c  c  c  c  c}
\tablecaption{Ratio of Effects}
\tablewidth{0pt}
\tablehead{ 
   \colhead{System Name} &
   \colhead{$\chi_{{\rm sec}}$} &
   \colhead{$\chi_{{\rm Kozai}}$} &
   \colhead{$\chi_{{\rm tide}}$} &
   \colhead{$\chi_{{\rm obl}}$} 
}
\renewcommand{\arraystretch}{1.5}
\startdata  
HD 11964 &     $ 1.02 $   & $6.43$    & $0.027$    & $0.0036$  \\
HD 38529 &     $ 0.79 $   & $6.11$    & $0.017$    & $0.006$   \\ 
HD 168443 &    $ 94.4 $   & $628.7$   & $0.0003$   & $0.0013$  \\
HD 190360 &    $ 0.14 $   & $1.05$    & $0.21$     & $0.0022$  \\ 
HD 108874 &    $ 1869.7 $ & $11068.1$ & $0.000003$ & $0.00029$ \\
\enddata
\tablecomments{ \label{Tabchi}
Values of the ratios, $\chi_{{\rm sec}} \equiv 
\dot{\omega}_{{\rm sec}}/\dot{\omega}_{{\rm GR}}$,
$\chi_{{\rm tide}} \equiv 
\dot{\omega}_{{\rm tide}}/\dot{\omega}_{{\rm GR}}$
and 
$\chi_{{\rm obl}} \equiv 
\dot{\omega}_{{\rm obl}}/\dot{\omega}_{{\rm GR}}$.  
The table values indicate that general relativity
may have an appreciable effect on the dynamical evolution
of HD 11964, HD 38529 and HD 190360, and that in most cases
for each of the 5 systems, effects from stellar oblateness 
and tidal effects are weak compared to those from general
relativity.  These values are computed from representative values
of the minimum masses and orbital parameters of each system.
}
\end{deluxetable}

\pagebreak

\begin{deluxetable}{ c  c  c  c  c  c  c  c  c  c}
\tabletypesize{\scriptsize}
\tablecaption{HD 11964 WITHOUT GENERAL RELATIVITY}
\tablewidth{0pt}
\tablehead{ 
   \colhead{$i_{{\rm rel}}$} &
   \colhead{Stable} &
   \colhead{$e_{b,med}$} &
   \colhead{$e_{b}$ curve} &
   \colhead{$\kappa_b$} &
   \colhead{$e_{c,med}$} &
   \colhead{$e_{c}$ curve} &
   \colhead{$\kappa_c$} &
   \colhead{$i_{r,med}$}
}
\renewcommand{\arraystretch}{1.5}
\startdata  
$0^{\circ}$ &  $100.0\%$ &  $ 0.28 $ & $[0.028,0.15,0.30,0.45,0.57] $ & $0.92$ & $0.068 $ & $[0.012,0.055,0.11,0.17,0.21] $ & $0.92$ & \nodata 
\\
$0^{\circ}-15^{\circ}$ &  $100.0\%$ &  $ 0.30 $ & $[0.032,0.16,0.32,0.48,0.61] $ & $0.90$ & $0.069 $ & $[0.012,0.055,0.11,0.17,0.21] $ & $0.92$ & $10.5^{\circ} $
\\
$15^{\circ}-30^{\circ}$ &  $100.0\%$ &  $ 0.31 $ & $[0.035,0.18,0.35,0.53,0.67] $ & $0.75$ & $0.071 $ & $[0.012,0.052,0.11,0.16,0.21] $ & $0.92$ & $23.9^{\circ} $ 
\\
$30^{\circ}-45^{\circ}$ &  $100.0\%$ &  $ 0.30 $ & $[0.035,0.18,0.36,0.55,0.69] $ & $0.47$ & $0.074$ & $[0.012,0.052,0.11,0.16,0.20] $ & $0.94$ & $38.7^{\circ} $ 
\\
$45^{\circ}-60^{\circ}$ &  $100.0\%$ &  $ 0.30 $ & $[0.045,0.22,0.44,0.66,0.83] $ & $0.32$ & $0.068 $ & $[0.012,0.052,0.11,0.16,0.21] $ & $0.91$ & $53.1^{\circ} $ 
\\
$60^{\circ}-75^{\circ}$ &  $48.8\%$ &  $ 0.28 $ & $[0.045,0.23,0.47,0.70,0.89] $ & $0.24$ & $0.074 $ & $[0.012,0.052,0.11,0.16,0.20] $ & $0.86$ & $64.4^{\circ} $ 
\\
$75^{\circ}-90^{\circ}$ &  $4.4\%$ &  $ 0.23 $ & $[0.048,0.24,0.48,0.72,0.92] $ & $0.17$ & $0.038 $ & $[0.0083,0.045,0.092,0.14,0.18] $ & $0.57$ & $83.7^{\circ} $ 
\\
$90^{\circ}-105^{\circ}$ &  $4.6\%$ &  $ 0.30 $ & $[0.078,0.27,0.50,0.74,0.93] $ & $0.19$ & $0.053 $ & $[0.0083,0.038,0.075,0.12,0.21] $ & $0.67$ & $95.8^{\circ} $
\\
$105^{\circ}-120^{\circ}$ &  $40.2\%$ &  $ 0.29$ & $[0.045,0.23,0.46,0.69,0.87] $ & $0.25$ & $0.067 $ & $[0.0083,0.048,0.098,0.15,0.24] $ & $0.87$ & $116.6^{\circ} $ 
\\
$120^{\circ}-135^{\circ}$ &  $99.6\%$ &  $ 0.30 $ & $[0.045,0.23,0.46,0.69,0.87] $ & $0.31$ & $0.067 $ & $[0.012,0.055,0.108,0.17,0.21] $ & $0.90$ & $127.1^{\circ} $
\\
$135^{\circ}-150^{\circ}$ &  $100.0\%$ &  $ 0.30 $ & $[0.038,0.20,0.40,0.59,0.75] $ & $0.45$ & $0.071 $ & $[0.012,0.052,0.11,0.16,0.21] $ & $0.93$ & $141.9^{\circ} $
\\
$150^{\circ}-165^{\circ}$ &  $100.0\%$ &  $ 0.29 $ & $[0.032,0.17,0.33,0.49,0.63] $ & $0.75$ & $0.066 $ & $[0.0083,0.045,0.092,0.14,0.18] $ & $0.91$ & $156.1^{\circ} $
\\
$165^{\circ}-180^{\circ}$ &  $100.0\%$ &  $ 0.28$ & $[0.035,0.17,0.35,0.52,0.66] $ & $0.89$ & $0.076 $ & $[0.012,0.055,0.11,0.16,0.21] $ & $0.93$ & $169.4^{\circ} $ 
\\
\enddata
%
\tablecomments{\small{  \label{Tabe11964}
Summary of stability, eccentricity evolution, and relative inclination
evolution for HD 11964 without the inclusion of GR arranged by bins 
of relative inclination (Column 1).  
Column 2 displays the percent of stable systems out of 500, 
Columns 3, 6 and 9 display median starting values, and Columns 4, and 7 
display the $5$, $25$, $50$, $75$ and $95$ percentiles of the 
time-averaged eccentricity across all stable integrations within
the given initial relative inclinations.  Columns 
5 and 8 display the mean ratios of the minimum vs. maximum eccentricities
obtained throughout the simulations.
}}
\end{deluxetable}

\pagebreak

\begin{deluxetable}{ c  c  c  c  c  c  c  c  c  c}
\tabletypesize{\scriptsize}
\tablecaption{HD 11964 WITH GENERAL RELATIVITY}
\tablewidth{0pt}
\tablehead{ 
   \colhead{$i_{{\rm rel}}$} &
   \colhead{Stable} &
   \colhead{$e_{b,med}$} &
   \colhead{$e_{b}$ curve} &
   \colhead{$\kappa_b$} &
   \colhead{$e_{c,med}$} &
   \colhead{$e_{c}$ curve} &
   \colhead{$\kappa_c$} &
   \colhead{$i_{r,med}$} 
}
\renewcommand{\arraystretch}{1.5}
\startdata  
$0^{\circ}$ &  $100.0\%$ &  $ 0.28 $ & $[0.028,0.15,0.30,0.44,0.57] $ & $0.95$ & $0.068 $ & $[0.012,0.052,0.11,0.16,0.21] $ & $0.94$ & \nodata 
\\
$0^{\circ}-15^{\circ}$ &   $100.0\%$ &  $ 0.28 $ & $[0.028,0.15,0.30,0.45,0.57] $ & $0.93$ & $0.068 $ & $[0.012,0.052,0.11,0.16,0.21] $ & $0.93$ & $10.6^{\circ} $ 
\\
$15^{\circ}-30^{\circ}$ &  $100.0\%$ &  $ 0.28 $ & $[0.028,0.15,0.30,0.45,0.57] $ & $0.83$ & $0.068 $ & $[0.018,0.098,0.20,0.29,0.37] $ & $0.94$ & $23.3^{\circ} $
\\
$30^{\circ}-45^{\circ}$ &  $100.0\%$ &  $ 0.28 $ & $[0.035,0.18,0.36,0.54,0.69] $ & $0.61$ & $0.068 $ & $[0.012,0.055,0.11,0.16,0.21] $ & $0.94$ & $38.5^{\circ} $
\\
$45^{\circ}-60^{\circ}$ &  $100.0\%$ &  $ 0.28 $ & $[0.042,0.21,0.41,0.62,0.79] $ & $0.36$ & $0.068 $ & $[0.012,0.055,0.11,0.16,0.21] $ & $0.90$ & $52.4^{\circ} $
\\
$60^{\circ}-75^{\circ}$ &  $88.2\%$ &  $ 0.28 $ & $[0.045,0.23,0.46,0.68,0.87] $ & $0.27$ & $0.066 $ & $[0.012,0.052,0.11,0.16,0.21] $ & $0.86$ & $66.9^{\circ} $ 
\\
$75^{\circ}-90^{\circ}$ &  $27.8\%$ &  $ 0.34 $ & $[0.048,0.24,0.48,0.72,0.92] $ & $0.28$ & $0.064 $ & $[0.0083,0.048,0.098,0.15,0.21] $ & $0.83$ & $81.8^{\circ} $
\\
$90^{\circ}-105^{\circ}$ &  $23.6\%$ &  $ 0.33 $ & $[0.052,0.24,0.48,0.72,0.92] $ & $0.28$ & $0.068 $ & $[0.0083,0.048,0.098,0.15,0.19] $ & $0.84$ & $96.7^{\circ} $
\\
$105^{\circ}-120^{\circ}$ &  $83.6\%$ &  $ 0.27 $ & $[0.045,0.23,0.45,0.67,0.85] $ & $0.26$ & $0.068 $ & $[0.012,0.055,0.11,0.16,0.21] $ & $0.84$ & $113.3^{\circ} $ 
\\
$120^{\circ}-135^{\circ}$ &  $100.0\%$ &  $ 0.28 $ & $[0.042,0.21,0.41,0.62,0.78] $ & $0.34$ & $0.068 $ & $[0.012,0.055,0.11,0.17,0.21] $ & $0.90$ & $127.3^{\circ} $
\\
$135^{\circ}-150^{\circ}$ &  $100.0\%$ &  $ 0.28 $ & $[0.042,0.20,0.41,0.61,0.77] $ & $0.59$ & $0.068 $ & $[0.012,0.052,0.11,0.16,0.20] $ & $0.94$ & $141.5^{\circ} $
\\
$150^{\circ}-165^{\circ}$ &  $100.0\%$ &  $ 0.28 $ & $[0.038,0.19,0.38,0.57,0.73] $ & $0.83$ & $0.068 $ & $[0.012,0.058,0.12,0.18,0.22] $ & $0.94$ & $156.8^{\circ} $
\\
$165^{\circ}-180^{\circ}$ &  $100.0\%$ &  $ 0.28 $ & $[0.028,0.15,0.30,0.45,0.57] $ & $0.93$ & $0.068 $ & $[0.012,0.052,0.11,0.16,0.20] $ & $0.93$ & $169.1^{\circ} $
\\
\enddata
\tablecomments{\small{  \label{Tabe11964GR}
Same as Table \ref{Tabe11964} but incorporating the effects of GR.
}}
\end{deluxetable}

\pagebreak

\begin{deluxetable}{c}
\tablecaption{HD 38529 WITHOUT GENERAL RELATIVITY}
\tablehead{ 
   \colhead{ } 
}
\startdata
 Table 4 is located on www.dimitriveras.com  \\  
\enddata
\label{Tabe38529}
\end{deluxetable}

\begin{deluxetable}{c}
\tablecaption{HD 38529 WITH GENERAL RELATIVITY}
\tablehead{ 
   \colhead{ } 
}
\startdata
 Table 5 is located on www.dimitriveras.com  \\  
\enddata
\label{Tabe38529GR}
\end{deluxetable}

\begin{deluxetable}{c}
\tablecaption{HD 108874}
\tablehead{ 
   \colhead{ } 
}
\startdata
 Table 6 is located on www.dimitriveras.com  \\  
\enddata
\label{Tabe108874}
\end{deluxetable}

\begin{deluxetable}{c}
\tablecaption{HD 168443}
\tablehead{ 
   \colhead{ } 
}
\startdata
 Table 7 is located on www.dimitriveras.com  \\  
\enddata
\label{Tabe168443}
\end{deluxetable}

\begin{deluxetable}{c}
\tablecaption{HD 190360 WITHOUT GENERAL RELATIVITY}
\tablehead{ 
   \colhead{ } 
}
\startdata
 Table 8 is located on www.dimitriveras.com  \\  
\enddata
\label{Tabe190360}
\end{deluxetable}

\begin{deluxetable}{c}
\tablecaption{HD 190360 WITH GENERAL RELATIVITY}
\tablehead{ 
   \colhead{ } 
}
\startdata
 Table 9 is located on www.dimitriveras.com  \\  
\enddata
\label{Tabe190360GR}
\end{deluxetable}

\begin{deluxetable}{ c  c  c  c  c}
\tabletypesize{\scriptsize}
\tablecaption{HD 11964 (WITH GENERAL RELATIVITY)}
\tablewidth{0pt}
\tablehead{ 
   \colhead{$i_{{\rm rel}}$} &
   \colhead{aligned} & 
   \colhead{antialigned} &
   \colhead{other} &
   \colhead{RMS $\Delta\varpi'$} 
}
\renewcommand{\arraystretch}{1.5}
\startdata  
$0^{\circ}$ & $68.6\% \ (68.4\%) \ [59.8\%]$  & $26.8\% \ (30.8\%) \ [6.2\%]$  & $4.6\% \ (0.8\%) \ [34.0\%]$  & $91.4^{\circ} \pm 5.0^{\circ} \ (95.7^{\circ} \pm 4.9^{\circ}) \ [95.2^{\circ} \pm 5.5^{\circ}]$
\\
$0^{\circ}-15^{\circ}$ & $65.8\% \ (66.8\%) \ [60.6\%]$  & $31.0\% \ (32.4\%) \ [8.8\%]$  & $3.2\% \ (0.8\%) \ [30.6\%]$  & $91.8^{\circ} \pm 4.3^{\circ} \ (94.9^{\circ} \pm 4.9^{\circ}) \ [96.3^{\circ} \pm 5.4^{\circ}]$
\\
$15^{\circ}-30^{\circ}$ & $60.6\% \ (66.2\%) \ [54.4\%]$  & $34.4\% \ (32.6\%) \ [10.2\%]$  & $5.0\% \ (1.2\%) \ [35.4\%]$  & $91.3^{\circ} \pm 6.5^{\circ} \ (93.8^{\circ} \pm 4.4^{\circ}) \ [96.6^{\circ} \pm 7.1^{\circ}]$
\\
$30^{\circ}-45^{\circ}$ & $39.6\% \ (57.0\%) \ [41.2\%]$  & $21.6\% \ (39.0\%) \ [13.2\%]$  & $38.8\% \ (4.0\%) \ [45.6\%]$  & $63.6^{\circ} \pm 25.4^{\circ} \ (91.5^{\circ} \pm 8.9^{\circ}) \ [70.9^{\circ} \pm 24.2^{\circ}]$
\\
$45^{\circ}-60^{\circ}$ & $37.6\% \ (51.2\%) \ [19.0\%]$  & $30.4\% \ (30.4\%) \ [6.6\%]$  & $32.0\% \ (18.4\%) \ [74.4\%]$  & $64.3^{\circ} \pm 25.0^{\circ} \ (78.8^{\circ} \pm 20.2^{\circ}) \ [65.4^{\circ} \pm 26.7^{\circ}]$
\\
$60^{\circ}-75^{\circ}$ & $50.0\% \ (54.2\%) \ [15.6\%]$  & $34.0\% \ (35.4\%) \ [9.0\%]$  & $16.0\% \ (10.4\%) \ [75.4\%]$  & $74.4^{\circ} \pm 21.5^{\circ} \ (80.3^{\circ} \pm 16.6^{\circ}) \ [66.7^{\circ} \pm 26.4^{\circ}]$
\\
$75^{\circ}-90^{\circ}$ & $50.0\% \ (54.7\%) \ [36.4\%]$  & $27.3\% \ (41.7\%) \ [13.6\%]$  & $22.7\% \ (3.6\%) \ [50.0\%]$  & $81.2^{\circ} \pm 12.6^{\circ} \ (84.0^{\circ} \pm 11.4^{\circ}) \ [75.7^{\circ} \pm 19.9^{\circ}]$
\\
$90^{\circ}-105^{\circ}$ & $60.9\% \ (40.7\%) \ [13.0\%]$  & $21.7\% \ (55.9\%) \ [21.7\%]$  & $17.4\% \ (3.4\%) \ [65.2\%]$  & $75.4^{\circ} \pm 20.2^{\circ} \ (83.7^{\circ} \pm 11.8^{\circ}) \ [70.0^{\circ} \pm 22.2^{\circ}]$
\\
$105^{\circ}-120^{\circ}$ & $42.3\% \ (40.0\%) \ [17.3\%]$  & $37.8\% \ (47.4\%) \ [8.4\%]$  & $19.9\% \ (12.7\%) \ [74.3\%]$  & $75.1^{\circ} \pm 21.5^{\circ} \ (78.5^{\circ} \pm 17.1^{\circ}) \ [69.8^{\circ} \pm 25.8^{\circ}]$
\\
$120^{\circ}-135^{\circ}$ & $31.3\% \ (35.0\%) \ [19.2\%]$  & $34.9\% \ (44.6\%) \ [8.0\%]$  & $33.7\% \ (20.4\%) \ [72.7\%]$  & $62.7^{\circ} \pm 27.4^{\circ} \ (77.7^{\circ} \pm 21.7^{\circ}) \ [64.9^{\circ} \pm 26.9^{\circ}]$
\\
$135^{\circ}-150^{\circ}$ & $34.2\% \ (55.0\%) \ [34.0\%]$  & $25.4\% \ (39.8\%) \ [15.2\%]$  & $40.4\% \ (5.2\%) \ [50.8\%]$  & $64.6^{\circ} \pm 25.4^{\circ} \ (91.1^{\circ} \pm 9.8^{\circ}) \ [69.9^{\circ} \pm 24.6^{\circ}]$
\\
$150^{\circ}-165^{\circ}$ & $60.8\% \ (61.4\%) \ [48.6\%]$  & $32.8\% \ (36.6\%) \ [12.6\%]$  & $6.4\% \ (2.0\%) \ [38.8\%]$  & $92.4^{\circ} \pm 5.8^{\circ} \ (94.4^{\circ} \pm 4.4^{\circ}) \ [95.8^{\circ} \pm 7.8^{\circ}]$
\\
$165^{\circ}-180^{\circ}$ & $57.0\% \ (78.8\%) \ [52.0\%]$  & $34.6\% \ (20.6\%) \ [11.2\%]$  & $8.4\% \ (0.6\%) \ [36.8\%]$  & $93.0^{\circ} \pm 5.6^{\circ} \ (95.0^{\circ} \pm 4.8^{\circ}) \ [96.5^{\circ} \pm 7.1^{\circ}]$
\\
\enddata
\tablecomments{\small{
\label{Tablib11964}
Summary of libration of $\Delta\varpi'$ about aligned configurations (Column 2),
antialigned configurations (Column 3) and asymmetric configurations 
(Column 4)  arranged by bins of relative inclination (Column 1) for HD 11964.
Column 5 displays the root mean squared (RMS) libration and 
its standard deviation across different initial conditions about the most 
prevalent libration center.  Values in parenthesis are computed when GR 
is included in the simulations, and values in square brackets refer 
to the libration of $\omega_{{\rm in}}$ measured with respect to the invariable plane.
}}
\end{deluxetable}

\begin{deluxetable}{c}
\tablecaption{HD 38529 WITH GENERAL RELATIVITY}
\tablehead{ 
   \colhead{ } 
}
\startdata
 Table 11 is located on www.dimitriveras.com  \\  
\enddata
\label{Tablib38529}
\end{deluxetable}

\begin{deluxetable}{c}
\tablecaption{HD 108874}
\tablehead{ 
   \colhead{ } 
}
\startdata
 Table 12 is located on www.dimitriveras.com  \\  
\enddata
\label{Tablib108874}
\end{deluxetable}

\begin{deluxetable}{c}
\tablecaption{HD 168443}
\tablehead{ 
   \colhead{ } 
}
\startdata
 Table 13 is located on www.dimitriveras.com  \\  
\enddata
\label{Tablib168443}
\end{deluxetable}

\begin{deluxetable}{c}
\tablecaption{HD 190360}
\tablehead{ 
   \colhead{ } 
}
\startdata
 Table 14 is located on www.dimitriveras.com  \\  
\enddata
\label{Tablib190360}
\end{deluxetable}

\pagebreak

\clearpage

\pagebreak

\clearpage

\pagebreak

\clearpage

\pagebreak

\clearpage

\end{document}